\documentclass[10pt,journal,compsoc]{IEEEtran}
\usepackage{hyperref}
\usepackage{comment}
\usepackage{amsmath}
\usepackage{newtxmath}
\usepackage{tikz}
\usetikzlibrary{quantikz}
\AtBeginDocument{%
  \providecommand\BibTeX{{%
    \normalfont B\kern-0.5em{\scshape i\kern-0.25em b}\kern-0.8em\TeX}}}
\usepackage{braket}
\usepackage[shortlabels]{enumitem}

\ifCLASSOPTIONcompsoc
  \usepackage[nocompress]{cite}
\else
  \usepackage{cite}
\fi
\ifCLASSINFOpdf
\else
\fi
\begin{document}
\title{Intermediate Qutrit-based Improved Quantum Arithmetic Operations with Application on Financial Derivative Pricing}
    % one-size-fits-all approach a ` prospect. For instance, compsoc 

\author{Amit Saha,
        Turbasu Chatterjee,
        Anupam Chattopadhyay,~\IEEEmembership{Senior~Member,~IEEE,}
        and~Amlan~Chakrabarti,~\IEEEmembership{Senior~Member,~IEEE}% <-this % stops a space
\IEEEcompsocitemizethanks{\IEEEcompsocthanksitem A. Saha is with ATOS, Pune, India\\
E-mail: abamitsaha@gmail.com
\IEEEcompsocthanksitem A. Saha, T. Chatterjee and A. Chakrabarti are with A. K. Choudhury School of Information Technology, University of Calcutta, India 
\IEEEcompsocthanksitem A. Chattopadhyay is with School of Computer Science and Engineering, Nanyang Technological University, Singapore.}}

\IEEEtitleabstractindextext{%
\begin{abstract}
In some quantum algorithms, arithmetic operations are of utmost importance for resource estimation. In binary quantum systems, some efficient implementation of arithmetic operations like, addition/subtraction, multiplication/division, square root, exponential and arcsine etc. have been realized, where resources are reported as a number of Toffoli gates or T gates with ancilla. Recently it has been demonstrated that intermediate qutrits can be used in place of ancilla, allowing us to operate efficiently in the ancilla-free frontier zone. In this article, we have incorporated intermediate qutrit approach to realize efficient implementation of all the quantum arithmetic operations mentioned above with respect to gate count and circuit-depth without T gate and ancilla. Our resource estimates with intermediate qutrits could guide future research aimed at lowering costs considering arithmetic operations for computational problems. As an application of computational problems, related to finance, are poised to reap the benefit of quantum computers, in which quantum arithmetic circuits are going to play an important role. In particular, quantum arithmetic circuits of arcsine and square root are necessary for path loading using the re-parameterization method, as well as the payoff calculation for derivative pricing. Hence, the improvements are studied in the context of the core arithmetic circuits as well as the complete application of derivative pricing. Since our intermediate qutrit approach requires to access higher energy levels, making the design prone to errors, nevertheless, we show that the percentage decrease in the probability of error is significant owing to the fact that we achieve circuit robustness compared to qubit-only works.

\end{abstract}

}

\maketitle

% is not selected <OR> if conference mode is selected - because compsoc
% conference papers position the abstract like regular (non-compsoc)
\IEEEdisplaynontitleabstractindextext

% \fi
\IEEEpeerreviewmaketitle

%\ifCLASSOPTIONcompsoc
%\IEEEraisesectionheading{\section{Introduction}\label{sec:introduction}}
%\else
\section{Introduction}

%\textcolor{red}{Here, add a discussion about - first, why financial derivative pricing is computationally demanding, and second, growing prowess of quantum computing to address it.}

\IEEEPARstart{Q}{uantum} computers are expected to outperform classical computers at certain computational tasks with an asymptotic advantage \cite{chuang}. While quantum algorithms for computationally difficult tasks have favourable asymptotic improvement, specific run-time estimates are frequently lacking due to the dearth of efficient implementation of quantum arithmetic operations that are employed as functions in quantum algorithms \cite{hner2018optimizing}. We have tried to address this issue by designing advanced arithmetic circuits with the adoption of higher energy level of quantum state. As an application, we consider financial derivative pricing problem to demonstrate our efficient approach of synthesizing arithmetic operations in quantum domain. 

In this article, we shall be focusing on the usage of quantum algorithms valuation or the pricing of these financial derivatives. Recently, it has been shown that quantum approaches to derivative pricing might exhibit quantum advantage with bounded error \cite{PhysRevA.98.022321}. The pricing methodology with a strong focus on the implementation of the algorithms in a gate-based quantum computer has also been addressed in \cite{Martin_2021, Zoufal_2019}. According to the most recent work on the decomposition of the solution of derivative pricing presented in \cite{Chakrabarti_2021}, at the granular level, one has to take into account the basic building blocks of the algorithm: The arithmetic operations, which encompass addition, subtraction, multiplication, square root, exponentiation, and the arcsine. An inefficient implementation of these blocks can employ colossal noise in the system, hence, derivative pricing algorithm may suffer with inappropriate outcome. 

%\textcolor{red}{Here, add a few more works in the reference that studies quantum derivative pricing. only reference to chakrabarti appears odd, as if the problem is not relevant enough}

The goal of this paper is to disintegrate the quantum derivative pricing algorithm presented in \cite{Chakrabarti_2021} into the arithmetic operations and put more efficient algorithms for the same at its foundation. In doing so, the methods prescribed in this paper do hope to carry out an asymptotic reduction of the complexity prescribed in the paper and improve the error bounds of the algorithm in its state-of-the-art form by decomposing Toffoli gates with the adoption of intermediate qutrit approach \cite{Gokhale_2019}. All the generalized qutrit gates \cite{Di11, Wang_2020} have already been successfully implemented on superconducting \cite{PhysRevA.76.042319}, trapped ion \cite{qutrit} and photonic systems \cite{Gao_2020} indicating it is possible to consider higher level systems apart from qubit-only systems. The Toffoli gate decomposition technique necessitates access to a higher energy level, rendering the design vulnerable to errors. Nonetheless, we demonstrate that the percentage reduction in the probability of error is significant because we reduced both gate count and circuit depth compared to previous work \cite{Chakrabarti_2021}.

Our specific contributions in a nutshell are as follows:

\begin{itemize}
    \item We employ intermediate qutrits to address basic quantum arithmetic operations  and  estimate the resources to show that our approach supersedes the current techniques on the basis of circuit depth and circuit robustness.
    \item Since arithmetic operations are an integral aspect of derivative pricing, hence we exhibit this as an example for application efficiently.
    \item Our numerical analysis establishes that adoption of intermediate qutrit for Toffoli decomposition is sublimer than the existing works by obtaining a percentage increase in the probability of success.
    %by $\simeq 40\%$ for Toffoli count 30 in a circuit.

\end{itemize}

The rest of this article is organized as follows: Section~\ref{sec:derivative-pricing} exhibits relevant preliminaries about
Derivative pricing with arithmetic operations. Section \ref{intermediate} demonstrates our intermediate qutrit approach to quantum arithmetic operations, and Section \ref{resource} analysis the resources required for proposed quantum arithmetic operations. The performance of the Toffoli decomposition with intermediate qutrit under various types of noise is presented in Section \ref{error}. Section \ref{conclusion} captures our concluding remarks.

\section{Derivative Pricing}~\label{sec:derivative-pricing}
In finance, the term `derivative' refers to a contract that derives its pricing from an underlying financial entity, such as, asset, index and interest rate. A derivative contract is typically valid until its expiration date and is issued between an issuer and a holder. History tells us that these derivatives date way back to the time of Aristotle~\cite{crawford_sen_1996}, and ever since then, these financial instruments saw a massive boom in trade during the last fifty years and have developed into complex financial instruments. The complexity of pricing a derivative in the modern market scenario is due to the nature of the underlying, on which they are being valued. Moreover, these financial instruments can be subject to a broad spectrum of usage, e.g., valuation, arbitrage, speculation and hedging. Calculating the fair value through derivative pricing can be difficult due to the stochastic nature of the parameters on which they are defined. Hence, many numerical approaches for derivative pricing are frequently used, with Monte Carlo being one of the most prominent due to its versatility and ability to handle stochastic parameters generically \cite{BOYLE1977323}. Monte Carlo methods, despite their appealing qualities in derivative pricing, typically demand a lot of computer resources to generate reliable option price estimates, especially for complex options. It is evident from the literature  that a quantum computer may provide novel ways to solve computationally intensive problems by leveraging quantum mechanical laws \cite{chuang}. More specifically, there have been multiple approaches to derivative pricing on the quantum frontier. We, however, focus our paper on the works of Chakrabarti et. al. \cite{Chakrabarti_2021}, wherein we come across rigorous resource estimates for executing derivative pricing on a quantum computer.

%This paper serves as a benchmark for the qubit count, T-depth, and the errors when implementing an algorithm for the valuation of a particular type of financial derivative: Options. The method outlined in \cite{Chakrabarti_2021}  uses a Quantum Amplitude Estimation as a primary block to evaluate the price of an option. 
The price of an underlying asset(s) is generally modeled as a stochastic process under the pretext of "no arbitrage," which assumes that an underlying asset(s) may not be priced differently in different markets, such that, there are no profits from trading between markets. The Black-Scholes method is used to model the evolution of the value of a financial asset as a geometric Brownian motion. Therefore, let $\vec{S^t} \in \mathbb{R}^{d}_+$ be a vector at time $t$ for $d$ underlyings. Let $(\vec{S^0}, \vec{S^1} \dots \vec{S^T}) = \omega \in \bar{\Omega}$ be a path of a discrete-time multivariate stochastic process describing the values of the underlying assets. The probability density function is denoted by $\bar{p}(\bar{\omega})$. Let $\bar{f}(\bar{\omega}) = f(\vec{S^0}, \dots \vec{S^T}) \in \mathbb{R}$ be the discounted payoff of some derivative on those assets. In order to evaluate the price of the derivative, we would calculate 
\begin{equation}
\mathbb{E}(f) = \int_{\bar{\omega} \in \bar{\Omega}}\bar{p}(\bar{\omega})\bar{f}(\bar{\omega}) d\bar{\omega}.
\end{equation}

The pricing for a given derivative can be calculated analytically using a variety of methods, notably the Black-Scholes model for \textit{path-independent} evaluations and Monte-Carlo methods for \textit{path-dependent} evaluations, the likes of autocallables and Target redemption forwards (TARFs). Computationally, it has been noted that the path-independent derivative is less intensive, while path-dependent evaluations still remain to be more intensive. Using Monte-Carlo methods classically, given $M$ paths, the derivative pricing accuracy converges as $O(1/\sqrt{M})$. This bound was improved upon by Montenaro and was found to be $O(1/M)$ in the quantum realm using quantum algorithms based on amplitude estimation, thus paving the way for investigations in derivative pricing, risk analysis and overall, quantum finance. 

In order to price the derivatives on a discrete variable quantum computer (DVQC), the paths need to be discretized and then mapped onto quantum states. Therefore the expectation now looks like 
\begin{equation}
\mathbb{E}(f) = \sum_{{\omega} \in  {\Omega}}{p}({\omega}){f}({\omega}).
\end{equation}

For path-dependent derivatives, the paths are generated using the underlying stochastic process and then the expected payoff is calculated using the estimator:

\begin{equation}
\mathbb{E}(f) \approx \frac{1}{N}\sum_{\omega = 1}^{M} f(\omega),
\end{equation}
which converges to the true expected value with error $\epsilon = O(M^{-1/2})$ by the central limit theorem. This convergence has been shown to exhibit quadratic speedup to $\epsilon = O(M^{-1})$ when using quantum amplitude estimation \cite{Brassard_2002} for Monte-Carlo processes \cite{Montanaro_2015, PhysRevA.98.022321, Stamatopoulos_2020}. Quantum amplitude estimation takes a unitary operator $\mathcal{A}$, acting on $n+1$ qubits such that, 
\begin{equation}
\mathcal{A}|0\rangle_{n+1} = \sqrt{1-a}|\psi_0\rangle_n|0\rangle + \sqrt{a}|\psi_1 \rangle_{n} |1\rangle,
\end{equation}
where the parameter $a$ is unknown and the final qubit acts as a label and is used to distinguish the $|\psi_0\rangle$ states from the $|\psi_1\rangle$ states.

The goal of the quantum amplitude estimation algorithm is to estimate the unknown parameter $a$. This is done by repeatedly applying the Grover operator $\mathcal{Q} = \mathcal{A} S_{0} \mathcal{A}^{\dagger}S_{\psi_0},$ where $S_0 = \mathbb{I} - 2|0 \rangle_{n+1}\langle 0 |_{n+1} $ and $S_{\phi_0} = \mathbb{I} - 2|\psi_0\rangle_n |0\rangle\langle0|\langle \psi_0 |_n$ are reflection operators. The parameter $a$ can be determined with an accuracy of $O(M^{-1})$ using quantum phase estimation and quantum Fourier transform \cite{chuang}. However, due to the use of the expensive quantum Fourier transform, this method has undergone development \cite{Suzuki2020, Aaronson_2020, Grinko2021, Nakaji_2020, Tanaka_2021, https://doi.org/10.48550/arxiv.2012.03348} and the Iterative Quantum Amplitude Estimation (IQAE) \cite{Grinko2021}, which has been regarded as the most efficient methods of the quantum amplitude estimation  has seen its use in \cite{Chakrabarti_2021}.

The basic approach to the derivative pricing algorithm consists of the following steps: 

\begin{enumerate}
    \item Apply some operator $\mathcal{P}$ to the state $|0 \rangle$, such that $$\mathcal{P} |0 \rangle = \sum_\omega \sqrt{p(\omega} |\omega\rangle$$
    
    \item Calculate $\delta(\omega) = \arcsin \sqrt{\tilde{f}(\omega)}$, such that $$\sum_\omega \sqrt{p(\omega)}|\omega\rangle |\delta(\omega) \rangle$$
    
    \item Introduce an ancilla qubit and use controlled rotations to rotate the value of $\tilde{f(\omega)}$ register into its amplitude, given by $$\sum_\omega \sqrt{p(\omega)(1-{\tilde f(\omega))}}|\omega\rangle|0\rangle + \sum_\omega \sqrt{p(\omega)({\tilde f(\omega))}}|\omega\rangle|1\rangle$$
    
    \item Use quantum amplitude estimation and extract the probability of the ancilla being $|1\rangle$
\end{enumerate}

The output yields the discretized expected payoff, $\mathbb{E}(\tilde f) = \sum_\omega p(\omega)\tilde f(\omega)$, subsequently rescaled to yield $\mathbb{E}(f) = (f_{max} - f_{min})\mathbb{E}(\tilde f) + f_{min}$.

In order for this algorithm to have a quantum advantage, one needs to take into account, the costs for path loading for the payoff calculation. The loading of general probability distributions have been proven to be exponentially hard. Therefore, one needs to find an efficient path loading operator $\mathcal{P}$. The Grover-Rudolph method is an efficient method for loading distributions which are integrable. This however is not useful for applications in derivative pricing as the probability distributions herein are not integrable and the paths are calculated using quantum Monte Carlo methodologies. Here, the author of \cite{Chakrabarti_2021} uses a variational method, termed as the Re-parameterization method to tackle the problem of path loading.

The Re-parameterization method works by observing that probability distributions may be loaded by preparing, in parallel, several standard normal distributions, and applying affine transformations to shape them into the desired distribution with the required mean and standard deviation. The steps for the Re-parameterization method can be outlined as follows: 

\begin{enumerate}
    \item Apply $dT$ Gaussian operators $\mathcal{G}$, to $ndT$ qubits, thereby constructing: 
    $$\bigotimes_{t=1}^T \bigotimes_{j=1}^d \mathcal{G}|0\rangle_n = \sum_{\omega_{\bar R}}\sqrt{p(\omega_{\bar R}})|\omega_{\bar R}\rangle_{ndT},$$
    where $\omega_{\bar R}$ runs over all $2^{ndT}$ different realizations of the multivariate standard Gaussian, and $p(\omega_{\bar R})$ are the corresponding probabilities. 
    
    \item Given $\Sigma = LL^T$ is the Cholesky decomposition of the covariance matrix, perform the affine transformation $\vec{R}^t = \vec{\mu}^t + L^T \vec{\bar{R}}^T$ to adjust the mean and standard deviation of each Gaussian. These will be used to calculate the asset prices from the log-returns given by $$S_j^{t'} = S_j^{t=0}e^{\mu_j t' + \sum_{i=0}^{d} L_{ji}^T \bar{R_i}^{t'}} = e^{ln S_j^{t=0} + \mu_j t' + \sum_{i=0}^{d-1} L_{ji}^T \bar{R_i}^{t'}}$$
    
    \item If the payoffs can be calculated directly from the log-returns, then the quantity $\delta(\omega_R) = arcsine \sqrt{\tilde f (\omega_R)}$ can be calculated in the quantum register as follows: 
    $$ \sum_{\omega_R}\sqrt{p(\omega_R)}|\omega_R\rangle|\delta(\omega_R) \rangle,$$
    where $\omega_R$ are the paths and $p(\omega_R)$ are the corresponding probabilities. If the payoff is defined in terms of prices and not just log-returns, then price path $\omega$ is calculated for each asset using $$\vec{S}^t = \vec{S}^0 e^{\sum_{j=1}^t \vec{R}^j}.$$
    
    This can be done in parallel for each asset.
    
    \item An ancilla qubit is introduced and the value of $\tilde{f}(\omega_R)$ is rotated into its amplitude: 
    $$\sum_\omega \sqrt{p(\omega)(1-{\tilde f(\omega))}}|\omega\rangle|0\rangle + \sum_\omega \sqrt{p(\omega)({\tilde f(\omega))}}|\omega\rangle|1\rangle$$
    
    \item Quantum amplitude estimation is used to extract the probability of the ancilla being $|1\rangle$
\end{enumerate}

The above routine can be visualized in figure \ref{fig:log_return}: 

\begin{figure}[h!]
    \centering
    \includegraphics[width=0.35\textwidth]{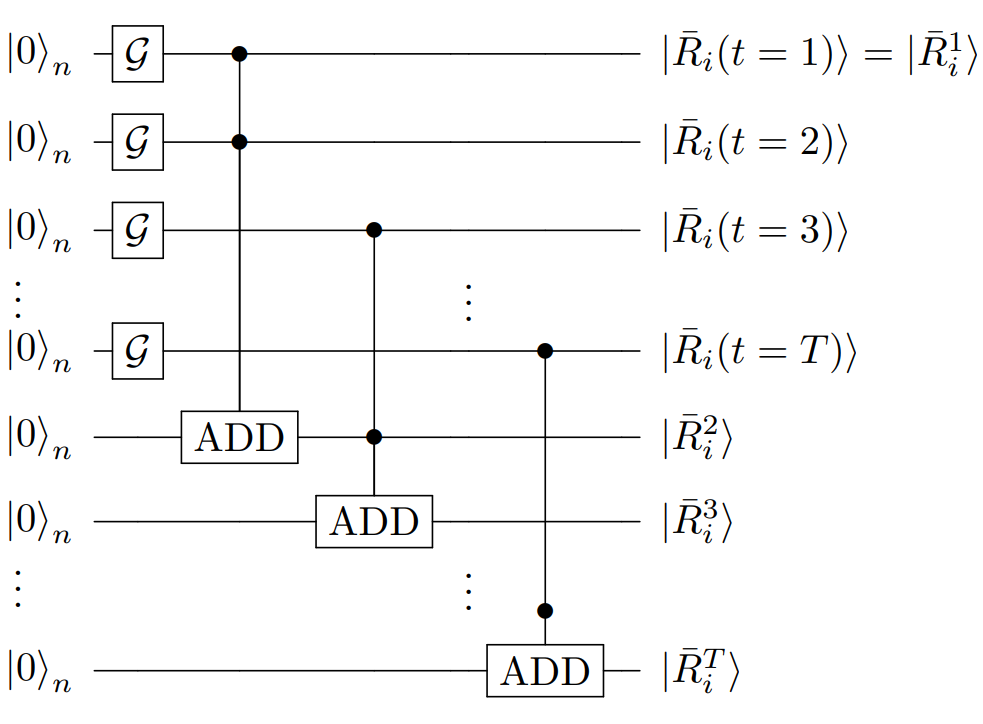}
    \caption{Circuit computing T registers containing cumulative log-returns $\bar{R^{t'}_i}$ for each timestep $t' \in [0,T]$. The $T$ Gaussian operators $\mathcal{G}$ are applied to the registers to generate Gaussian probability distributions at each timestep. The ADD operators are then used in series to transform $|x\rangle|y\rangle|0\rangle \mapsto |x\rangle|y\rangle|x+y\rangle$ \cite{Chakrabarti_2021}.}
    \label{fig:log_return}
\end{figure}

As the Gaussian path loading operator $\mathcal{G}$ takes the place of the operator $\mathcal{P}$, it is imperative to talk about the construction of $\mathcal{G}$. The authors of \cite{Chakrabarti_2021} use a Variational Quantum Eigensolver approach that uses a parameterized circuit, which in turn produce a parameterized quantum state $|\psi ( \lbrace \theta \rbrace) \rangle$, which approximately represents the target state $|\phi_0\rangle$. The parameters $\lbrace \theta \rbrace$ are updated using a classical optimizer that optimizes expectation value of a suitable cost function. This can be formulated by 
\begin{equation}
E_0 = \langle \psi_0 | H | \psi_0 \rangle,
\end{equation}
where energy expectation $E$, reaches its ground state $E_0$, when calculated on a target state. It is noted that the energy Hamiltonian of the quantum harmonic oscillator is given by: 
\begin{equation}
H = \frac{P^2}{2m} + \frac{m(X-x_0)^2}{2}, 
\end{equation}
furthermore, by clever observation, it is noted that this Hamiltonian has a ground state energy 
\begin{equation}
\phi_0 (x) = (\frac{m}{\pi})^{1/4} e^{-m(x-x_0)^2},
\end{equation}
which is a Gaussian function. Here $X$ is the position operator in real space, $P = -i\frac{d}{dx}$ is the momentum operator, $m$ is a parameter representing the variance of the Gaussian distribution, whereas $x_0$ is the center of the distribution. By setting $m = 1/(2\sigma^2)$, one can find the state $\psi_0 (x)$, such that $\psi_0^2(x) = \mathcal{N}(x_0, \sigma)$.

In order to calculate these on a quantum computer one must observe that $X^2$ is diagonal in the $Z$-basis, therefore it can be measure directly from the qubit register as $N_{counts}$ wave function collapses on measurement. The $P^2$ is diagonal on the momentum basis and therefore a centering Quantum Fourier Transform (QFT) is invoked after the state preparation block. Since the calculations are done on the discrete position space, the transformation $x_i = -w + i\Delta x$ is used, where $i=0,\cdots, 2^n-1$ and $\Delta x = 2w/2^n$. Without loss of generality, if the center is chosen to be at zero, the energy $E = E_{X^2} + E_{P^2}$ is calculated as follows: 
\begin{equation}
E_{X^2} = \frac{1}{N_{shots}}\sum^{\mathcal{N}}_{j=0} \frac{m}{2} N_{counts}(j)(j \times \Delta x - x_0)^2
\end{equation}
\begin{equation}
E_{P^2} = \frac{1}{N_{shots}}\sum^{\mathcal{N}}_{j=0} \frac{1}{2m} N_{counts}(j)(j \times \Delta p)^2
\end{equation}

where, $N_{shots}$ is the total number of circuit repetitions for the spacial and the momentum basis. Thus derivative pricing problem can be solved using re-parameterization method followed by payoff calculation efficiently with the help of arcsine and square root quantum operation. But there is no free lunch, hence, these methods have to deal with several errors which are discussed forthwith.

\subsubsection*{Error Analysis} The authors of \cite{Chakrabarti_2021} have characterized the errors induced in their approach as four main components: Truncation Error, Discretization Error, Amplitude Estimation Error and arithmetic error. First three are discussed here and the arithmetic error is discussed in the next subsection \ref{arith} for the sake of understanding.  

\begin{enumerate}[(a)]
    \item \textit{Truncation Error.} Given the fault-tolerant setting of the algorithm presented in \cite{Chakrabarti_2021}, to feasibly compute the infinitely many path integrals, the domain of integration is restricted to a range $[ B_l, B_u ]$, thereby leaving out a probability mass $\alpha$. Given an upper bound $f_{\delta} = f_{max}^{disc} - f_{min}^{disc}$ on the discounted payoff as calculated, and an upper bound $P_{max}$ on the density function at each step, a truncation error of $\epsilon_{trunc} = P^T_{max} f_\delta \alpha$ is incurred. For $dT$ different $n$-qubit registers that are $w$ standard deviations around the mean for each timestep, the truncation error is given by: 
    \begin{equation}
    \epsilon_{trunc} \leq 2dTe^{-w^2/2}
    \end{equation}
    
    \item \textit{Discretization Error.} This error arises when approximating the integral over finite grid of points and can be reduced by increasing the number of qubits $n$. Assuming $\beta$ is an upper bound on the second derivative of the integrand and given $n$ qubits, thereby discretizing the integral domain over $2^{ndT}$ cells, the discretization error $\epsilon_disc$ is given by: 
    \begin{equation}
    \epsilon_{disc} = \frac{\beta(B_u - B_l)^{dT+2}}{24 \cdot 2^{2n}}
    \end{equation}
    
    \item \textit{Amplitude Estimation Error.} An error of $\epsilon_{amp}$ is incurred by the Amplitude Estimation subroutine when repeating the state preparation and pricing procedure $1/\epsilon_{amp}$ times.
\end{enumerate}

A key observation that lays the very foundation of this paper is that the algorithm heavily depends on quantum arithmetic operations in its various aspects of its calculation in the discrete computational domain. It is therefore imperative, that these operations be probed, thereby paving the way for more efficient calculation methods in the quantum discrete arithmetic operation domain. These quantum arithmetic operations are further elaborated in the subsequent subsection. Since our approach of solving derivative pricing uses higher dimensional quantum systems, an additional error apart from these four has been inducted automatically, which is thoroughly described in Section \ref{error}.

\subsection{Quantum Arithmetic Operations}\label{arith}

\begin{figure*}[h!]
    \centering
    \includegraphics[width=0.8\textwidth]{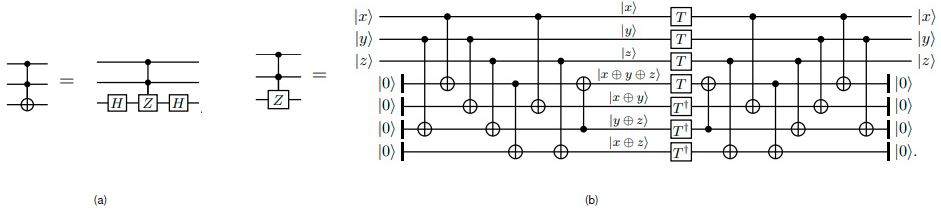}
    \caption{(a)Toffoli decomposition; (b)three-qubit phase gate decomposition }
    \label{fig:tof_selinger}
\end{figure*}

This section highlights the background
for common quantum arithmetic operations.
These operations are used in resource estimation and error analysis in derivative pricing as discussed in previous section.
In this section, resources are usually reported as Toffoli cost as we are working in the fault-tolerant setting. As per the state-of-the-art work, they also estimate the optimized T-depth of the
circuits by decomposing Toffoli gate in a Clifford + T gate set \cite{Selinger_2013}.  They decomposed the Toffoli gates with a T-depth
of one using four ancilla qubits as shown in Fig. \ref{fig:tof_selinger}  \cite{Selinger_2013}.  Combining Fig. \ref{fig:tof_selinger}(a) and \ref{fig:tof_selinger}(b), one can obtain a representation of the Toffoli gate
of T -depth 1 and overall depth 7. They have also tried to parallelize the resulting
circuits wherever possible to obtain the optimized depth.

The authors of \cite{Chakrabarti_2021} performed all the calculations in fixed-point arithmetic similarly to
\cite{hner2018optimizing}, which allows us to use the described quantum techniques for reversible function evaluation.
An $n$-bit representation of a number $x$ is

\begin{equation}
\label{eqn:fixed_point_repr}
x=\underbrace{x_{n-1}\cdots x_{n-p}}_p.\underbrace{x_{n-p-1} \cdots x_0}_{n-p},
\end{equation}
where $x_i \in {0,1}$ denotes the $i$-th bit of the binary representation of
$x$ and $p$ denotes the number of bits to the left of the binary decimal point.
The choice of $n$ and $p$ controls the error that one can allow in each calculation, we allow the same approximation for arithmetic error as in \cite{Chakrabarti_2021}. Let $Toffoli\_count$ and $T\_depth$ denote the number of Toffoli gates and the T-depth required to compute an arithmetic function as per the state-of-the-art work.

\paragraph*{Addition/Subtraction}
As per \cite{Chakrabarti_2021, draper2006logarithmic}, one can perform addition of two
$n$-qubit registers in place with a Toffoli cost, 

\begin{multline}
 Toffoli\_count_{add} =  10n-3w(n)-3w(n-1)-3\log_2n\\ - 3\log_2(n-1) -7 
\end{multline}
where $w(n)$ denotes the number of ones in the binary expansion of $n$,
and a Toffoli depth,

\begin{multline}
T\_depth_{add}=\lfloor \log_2(n)\rfloor + \lfloor\log_2(n-1)\rfloor + \lfloor\log_2\left(\frac{n}{3}\right)\rfloor \\+\lfloor\log_2\left(\frac{n-1}{3}\right)\rfloor + 8.
\end{multline}

For subtraction, the Toffoli count remains same as addition.

\paragraph*{Multiplication} For multiplication the authors follow the method, which uses the controlled addition circuit in \cite{takahashi2009quantum} and requires a Toffoli count,

\begin{equation}
Toffoli\_count_{\text{mul}}(n,p) = \frac{3}{2}n^2 + 3np + \frac{3}{2}n - 3p^2 + 3p.
\end{equation}
This method can also be used for division, where the Toffoli count remains same as multiplication. The fixed-point multiplication method's controlled additions necessitate ancilla qubits proportional to the register size, but the circuits support uncomputing the ancillas, allowing them to be reused for each consecutive addition that is not done in parallel. We may also parallelize each multiplication circuit by viewing each factor's register as $z \geq 1$ separate registers of size $n/z$, and each controlled addition for the $z$ subregisters can happen in parallel.
To aggregate the $z$ sub-results into the final result, $n \cdot (z-1)$ more qubits and $z-1$ additions are required.
$z=1$ indicates that no additional parallelization is used.
We can get a total T-depth cost of parallelized fixed-point multiplication by parallelizing the pairwise accumulation adds as well,

\begin{equation}
\label{eqn:t_depth_parallel_mult}
T\_depth_{\text{mul}}(n, z) = \lceil\frac{n}{z}\rceil \cdot (T\_depth_{add} + 6) + \lceil\log_2z\rceil \cdot T\_depth_{add}.
\end{equation}
$(\text{T}_{\text{add}} + 6)$ is the T-depth of a controlled addition discussed in the Addition/Subtraction section.

\paragraph*{Square Root} In \cite{Chakrabarti_2021}, they employ the square root algorithm from \cite{hner2018optimizing}, which has the Toffoli count,
\begin{equation}
Toffoli\_count_{\text{sq}}(n,p) = \frac{n^2}{2} + 3n - 4.
\end{equation}
The T-depth of this algorithm as reported by the authors is
\begin{equation}
    T\_depth_{\text{sq}}(n)= 5n + 3
\end{equation} 
where $2n+1$ qubits are required.

\paragraph*{Exponential} In \cite{Chakrabarti_2021}, the authors
developed a generic quantum technique that uses a parallel piecewise polynomial approximation to generate smooth classical functions as per \cite{Munoz2018}, which they used to estimate the computing resources for exponentials.
The procedure takes two parameters, $k$ and $M$, which regulate the piecewise approximation polynomial degree and the number of domain subintervals, respectively.
The total number of Toffolis is calculated as follows:

%\begin{equation}
\begin{multline}
Toffoli\_count_{\text{exp}}(n, p, k, M) = \frac{3}{2}n^2k + 3npk + \frac{7}{2}nk - 3p^2d \\+ 3pk - d + 2Md(4\lceil\log_2M\rceil - 8) + 4Mn.
\end{multline}
%\end{equation}

\paragraph*{Arcsine} 
The arcsine function is computed using the above exponential approach, which needs $k$ iterations of multiplication and addition, with $k$-degree polynomials employed for approximation.
It also necessitates $M$ comparison circuits between the $n$-qubit input register and a classical value for $M$ specified subintervals.
With the use of the comparator from \cite{draper2006logarithmic} with T-depth of
$2\lfloor\log_2(n-1)\rfloor+5$, the T-depth of a parallel polynomial evaluation circuit is
\begin{multline}
T\_depth_{\text{pp}}(n, z) = k\left(T\_depth_{\text{mul}}(n, z) + T\_depth_{\text{add}}\right) \\+ M(2\lfloor\log_2(n-1)\rfloor+5),
\end{multline}
%\end{equation}
where $z$ is the optional parallelization factor for the multiplication circuit introduced in the resource estimation above.

The total Toffoli count for computing $\ket{\arcsin\sqrt{x}}$ as per \cite{hner2018optimizing} is
\begin{multline}
Toffoli\_count_{\text{arcsq}}(n,p, k, M) = \\k\left(\frac{3}{2}n^2 + n(3p + \frac{7}{2}) -3(p-1)p -1\right) \\+ \frac{n^2}{2}  + 11n + 2Md(4\lceil\log_2M\rceil - 8) + 4Mn - 2.
\end{multline}

The T-depth for computing $\arcsin(\sqrt{x})$ of a number $x$ represented in a register of size $(n,p)$, calculated as akin to the exponential is
\begin{equation}
T\_depth_{\text{arcsq}}(n, p, z) = T\_depth_{\text{sq}}(n) + T\_depth_{\text{pp}}(n, z) + 8n + 6,
\end{equation}
where $T\_depth_{\text{sq}}(n)= 5n + 3$ is the T-depth for the square root algorithm from \cite{Munoz2018}. Next we portray our approach of intermediate qutrit on quantum arithmetic operations.

\section{Quantum Arithmetic Operations with Intermediate Qutrits}\label{intermediate}

This section first describes the Toffoli decomposition with the help of intermediate qutrit that we have adapted from \cite{Gokhale_2019}. Thenceforth, we discuss about one of the quantum arithmetic operations $i.e.,$ quantum adder with intermediate qutrits \cite{10.1007/s11128-007-0057-2, adderqudit}. Later we have exhibited quantum multiplier using intermediate qutrits with an example implementation. We have simulated this multiplier circuit on Google Colab platform \cite{Bisong2019} and the code is available at \href{https://github.com/LegacYFTw/NTU}{https://github.com/LegacYFTw/NTU}. These two operations, adder and multiplier, are the basic building blocks of other arithmetic operations like square root, exponential and arcsine, which are used in derivative pricing. Therefore, we vividly discuss about quantum adder and multiplier and their implementation further in this section.

\subsection{Toffoli Decomposition with Intermediate Qutrits}

\begin{figure}[h!]
    \centering
    \includegraphics[scale=.5]{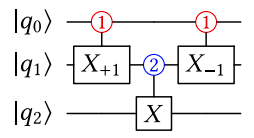}
    \caption{An example of Toffoli decomposition with intermediate qutrit, where input and output are qubits. The red controls activate on $\ket{1}$
and the blue controls activate on $\ket{2}$. The first gate temporarily elevates $q_1$ to $\ket{2}$ if both $q_0$ and $q_1$ were $\ket{1}$. $X$ operation is then only performed if $q_1$ is $\ket{2}$. The final gate acts as a miror of first gate and  restores $q_0$ and $q_1$ to their original state \cite{Gokhale_2019}}.
    \label{tof_qutrit}
\end{figure}

 In \cite{Gokhale_2019, 10.1145/3406309}, the authors showed that we may occupy the $\ket{2}$ state temporarily during computation, hence temporarily ternary. Maintaining binary input and output allows this circuit construction to be inserted into any pre-existing qubit-only circuits.  A Toffoli decomposition via qutrits has been portrayed in Fig. \ref{tof_qutrit} \cite{Gokhale_2019, 10.1145/3406309}. More specifically the goal is to carry out a NOT operation on the target qubit (third qubit) as long as the two control qubits, are both $\ket{1}$. First, a $\ket{1}$-controlled $X_{+1}$, where $+1$ is used to denote that the target qubit is incremented by $1 \ (\text{mod } 3)$, is performed on the first and the second qubits. This upgrades the second qubit to $\ket{2}$ if and only if the first and the second qubits were both $\ket{1}$. Then, a $\ket{2}$-controlled $X$ gate is applied to the target qubit. Therefore, $X$ is executed only when both the first and the second qubits were $\ket{1}$, as expected. The controls are reinstated to their original states by a $\ket{1}$-controlled $X_{-1}$ gate, which reverses the effect of the first gate. That the $\ket{2}$ state from ternary quantum systems can be used instead of ancilla to store temporary information, which is the most important aspect in this decomposition. Thus, to decompose Toffoli gate, 3 generalized ternary CNOT gates are sufficient with circuit depth 3. In fact, no T gate is required. 

\subsection{Quantum Adder with Intermediate Qutrits}

\begin{figure}
    \centering
    \includegraphics[scale=0.5]{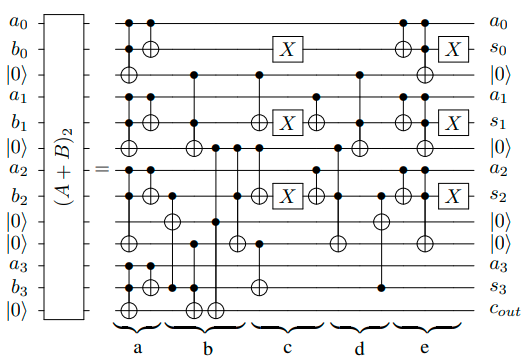}
    \caption{An example A+B quantum Adder with 4-bit A and 4-bit B \cite{adderqudit, draper2006logarithmic}}
    \label{fig:adder}
\end{figure}

We have now demonstrated quantum adder circuit from \cite{adderqudit, draper2006logarithmic} on two four-bit registers $A$ and $B$ with
a carry-out bit using ancilla as shown in Fig. \ref{fig:adder}. The sum $S$ is computed in-place on register B
while A is untouched and the ancilla are restored to $\ket{0}$. In \cite{adderqudit}, for the first time, the authors showed how this adder can be efficiently implemented with the help of intermediate qutrits with respect to gate count and circuit depth instead of conventional way of decomposition. The author achieves an advanced adder by decomposing  all the Toffoli gate presented in Fig. \ref{fig:adder} with the discussed intermediate qutrit method. This work provides us enough motivation to work upon others quantum operations with intermediate qutrit. Therefore, we present an example of implementing a multiplier with the adoption of intermediate qutrit for better understanding.

\begin{comment}
\begin{figure}
    \centering
    \includegraphics[scale=0.15]{Multiplier.png}
    \caption{Multiplier}
    \label{fig:multiplier}
\end{figure}
\end{comment}

\begin{comment}

\begin{figure}
    \centering
    \includegraphics[scale=0.028]{Square_root.png}
    \caption{Square Root}
    \label{fig:square}
\end{figure}

\end{comment}

\subsection{Quantum Multiplier with Intermediate Qutrits}

\begin{figure*}[!h]
    \centering
    \includegraphics[scale=0.4]{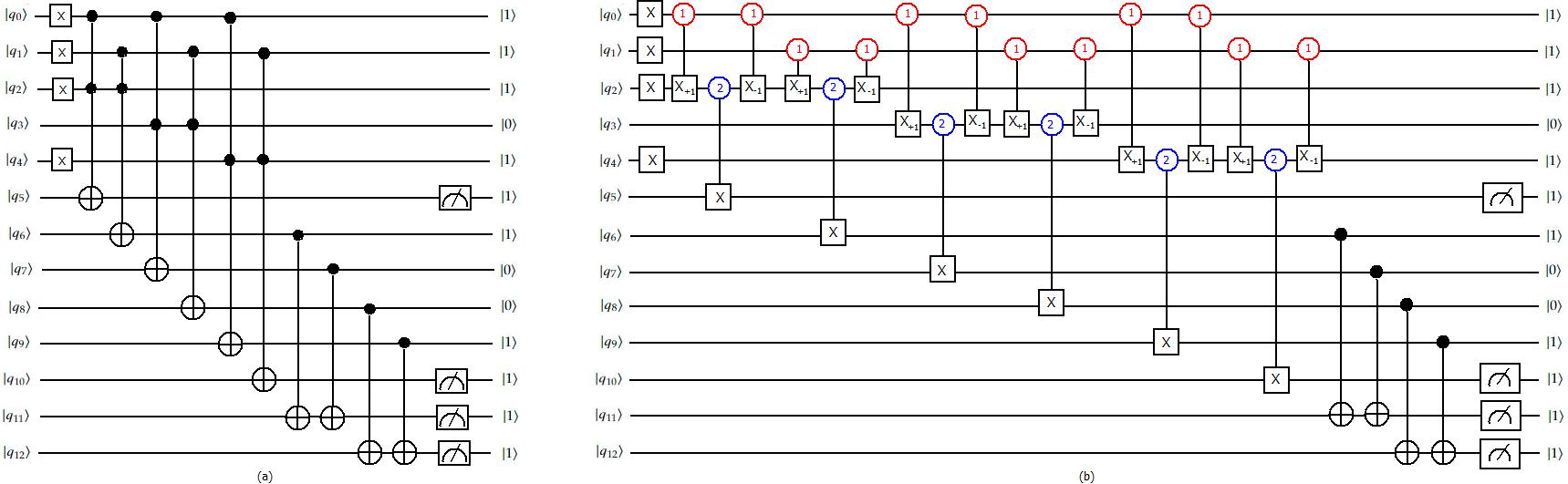}
    \caption{(a) Quantum Multiplier Circuit for the multiplication of $5 \times 3$; (b) Quantum Multiplier with Intermediate Qutrit for the multiplication of $5 \times 3$.}
    \label{fig:multiplier}
\end{figure*}

Let's put some light on quantum multiplier here in this subsection. We have considered multiplication of $5 \times 3$ as an example to illustrate our approach of designing quantum multiplier with intermediate qutrit. In Fig. \ref{fig:multiplier}(a), a multiplier circuit has been presented considering the mentioned example as per \cite{takahashi2009quantum}, in which all the qubits are initialized with $\ket{0}$. In this circuit, first five qubits ($q_0$ - $q_4$) are the input qubits, where first two qubits ($q_0$ and $q_1$) represent the number 3 by applying two NOT gates on them and other three qubits ($q_2$, $q_3$ and $q_4$) represent the number 5 by applying two NOT gates on qubits $q_2$ and $q_4$. Next, we perform multiply operation on these qubits by applying Toffoli gates and stores the value on ancilla qubits ($q_5$ - $q_{10}$). Now, we carry out addition by applying CNOT gates on ancilla qubits. Finally, to get the output of $5 \times 3$, we shall measure the qubits ($q_5$, $q_{10}$, $q_{11}$ and $q_{12}$). Further, all the Toffoli gates presented in Fig. \ref{fig:multiplier}(a) are decomposed using intermediate qutrit approach as shown in Fig. \ref{fig:multiplier}(b) to achieve asymptotic improvement of the circuit. Our numerical analysis also yields $5 \times 3 = 15$ perfectly. Since the circuits of quantum adder and multiplier are verified, the focus shifts to resource required to perform all the discussed quantum operations using intermediate qutrits.  

\begin{comment}

\begin{figure}[!h]
    \centering
    \includegraphics[scale=0.17]{multi_binary.eps}
    \caption{Quantum Multiplier Circuit for the multiplication of $5 \times 3$}
    \label{fig:multiplier}
\end{figure}

\begin{figure}[!h]
    \centering
    \includegraphics[scale=0.182]{Multiplier_qutrit.eps}
    \caption{Quantum Multiplier with Intermediate Qutrit for the multiplication of $5 \times 3$}
    \label{fig:multiplierqutrit}
\end{figure}

\end{comment}

\section{Resource Estimation and Analysis}\label{resource}

As shown in the proposed approach, we have synthesized the quantum arithmetic operations via intermediate qutrits, in which, three generalized ternary CNOT gates are used to decompose the Toffoli gate without any ancilla. Hence, no T gate has been used in any proposed circuit synthesis. Therefore, T-depth becomes zero in our proposed approach. The overall circuit depth for Toffoli decomposition has now become three as compared to seven. Since there are no T gates in our proposed circuits, the resource analysis is based on the count of generalized ternary CNOT gate, which is a Clifford gate. It also needs to be noted that the additional four ancilla qubits are not required to decompose Toffoli gate, otherwise the number of logical qubits remain same for the derivative pricing as per \cite{Chakrabarti_2021}, as no other optimization with respect to logical qubits is presented in this article. Hence, from now on we shall only discuss about gate count or circuit depth of derivative pricing problem.

\paragraph*{Addition/Subtraction}
As per our proposed approach, one can perform addition of two
$n$-qubit registers in place with a CNOT-cost, 

\begin{multline}
 CNOT\_count_{add} =  30n-9w(n)-9w(n-1)-9\log_2n \\- 9\log_2(n-1) -21 
\end{multline}
where $w(n)$ denotes the number of ones in the binary expansion of $n$. For subtraction, the CNOT count remains same as addition.

\paragraph*{Multiplication} For multiplication if one can follow our proposed method, then the CNOT count becomes,

\begin{equation}
CNOT\_count_{\text{mul}}(n,p) = \frac{9}{2}n^2 + 9np + \frac{9}{2}n - 9p^2 + 9p.
\end{equation}
This method can also be used for division, where the CNOT count remains same as multiplication.

\paragraph*{Square Root} If one can employ our proposed approach for the square root algorithm, then the CNOT count becomes,
\begin{equation}
CNOT\_count_{\text{sq}}(n,p) = \frac{3n^2}{2} + 9n - 12.
\end{equation}

\paragraph*{Exponential} As per our proposed approach,
the total number of ternary CNOT is given by

\begin{multline}
CNOT\_count_{\text{exp}}(n, p, k, M) = \frac{9}{2}n^2k + 9npk + \frac{21}{2}nk - 9p^2d \\+ 9pk - 3d + 6Md(4\lceil\log_2M\rceil - 8) + 12Mn.
\end{multline}

\paragraph*{Arcsine}

The total ternary CNOT count for computing $\ket{\arcsin\sqrt{x}}$ is
\begin{multline}
CNOT\_count_{\text{arcsq}}(n,p, k, M) = \\3k\left(\frac{3}{2}n^2 + n(3p + \frac{7}{2}) -3(p-1)p -1\right) + \frac{3n^2}{2} \\ + 33n + 6Md(4\lceil\log_2M\rceil - 8) + 12Mn - 26.
\end{multline}

\subsection{Comparative Analysis}

As discussed, the Toffoli decomposition with intermediate qutrit is novel not only for its depth optimization from 7 to 3 or gate count optimization from 25 to 3 as compared to \cite{Selinger_2013}, but also it does not require any T gates. Hence to execute required quantum arithmetic operations like arcsine and square root for derivative pricing, T-depth and T-cost become 0 for the proposed methodology as discussed in this article. Since the T gates are more expensive than other gates in terms of space and time cost due to their increased tolerance to noise errors, our method of implementing derivative pricing problem is superior to the method proposed in \cite{Chakrabarti_2021}. As there are no T gates in our methodology, the resources for all the arithmetic operations for derivative pricing are presented in the form of ternary CNOT gates, which is thoroughly discussed earlier in this section. It can be easily observed that the overall CNOT-cost for derivative pricing is much more improved as compared to \cite{Chakrabarti_2021}, since they are decomposed their Toffoli gate with 16 CNOT gates and as per our decomposition it requires only 3 CNOT gates. As a case study of derivative pricing problem in \cite{Chakrabarti_2021}, the authors have taken into consideration a basket autocallable (Auto) with 3 underlyings, 5 payment dates and a knock-in put option with 20 barrier dates. They found that the benchmark use case that they examined for re-parameterization method of derivative pricing, required T-cost of 12 billion, T-depth of 54 million and overall circuit depth at least 378 million. In our case, both of these T-cost and T-depth become 0, while the ternary CNOT-cost and overall circuit depth are 162 million respectively. These results evidently show that our method outperforms the run-time of the state-of-the-art algorithm for derivative pricing quite comprehensibly since gate count and circuit depth have been significantly improved. Although our circuit constructions have adopted higher dimensional gates, our used decomposition scales favorably in terms of asymptotically fewer gate errors and idle errors as our gate count and circuit depth is  asymptotically lower, which is elaborated upon in next section.

\section{Error Analysis of Toffoli decomposition with Intermediate qutrits}\label{error}

Decoherence, noisy gates, and other sorts of errors can occur in any finite-dimensional quantum system. It has been proven that using higher dimensional states other than binary systems causes the system to have more errors. The impact of these inaccuracies on the Toffoli gate decomposition has been investigated in this section. Although the introduction of qutrits increases error, the overall error probability of the decomposition is lower than that of the earlier decomposition \cite{Selinger_2013} since the number of ancilla qubits, gate count, and depth are reduced.

\subsection{Generic Error Model}
The conventional quantum error or noise model is for gate and relaxation error \cite{Gokhale_2019}, which can be expressed by the Kraus Operator formalism \cite{chuang}. If the density matrix representation of a (pure) quantum state is $\sigma = \ket{\Psi}{\bra{\Psi}}$, the evolution of this state for any channel is represented as the function $\mathcal{E}(\sigma)$:

\begin{align}
    \mathcal{E}\left(\sigma\right) = \mathcal{E}\left(\ket{\Psi}\bra{\Psi}\right) = \sum_i K_i \sigma K_i^\dagger
\end{align}
where $K_i$ are called the Kraus Operators, and $K_i^\dagger$ is the matrix conjugate-transpose of $K_i$, $\forall$ $i$. The Kraus operator formulation can also be used to represent the evolution of a state under a noise model. The Kraus operators, for example, are simply the Pauli matrices in the depolarization noise model.

\subsubsection{Gate Errors}
In a binary quantum system with only one-qubit and two-qubit gates, there are four possible error channels for a one-qubit gate, which can be expressed as products of the two Pauli matrices, a NOT gate, $X = \begin{pmatrix}
0 & 1 \\
1 & 0
\end{pmatrix}$ and a phase gate, $Z = \begin{pmatrix}
1 & 0 \\
0 & -1
\end{pmatrix}$. The possible error channels are: (i) no-error $X^0 Z^0 = I$, (ii) the phase flip which is the product $X^0 Z^1$, (iii) the bit flip which is $X^1 Z^0$ and (iv) the phase+bit flip channel given by $X^1 Z^1$. We can express this one-qubit gate error model in the Kraus operator formalism in the following manner:
\begin{align}
    \mathcal{E}(\sigma) = \sum_{j=0}^{1} \sum_{k=0}^{1} p_{jk} (X^j Z^k) \sigma (X^j Z^k)^{\dagger}
\end{align}
where $p_{jk}$ is the probability of the corresponding Kraus operator. 

%Let us assume all error terms have equal probabilities, i.e. $p_{jk} = p$ for $j,k \neq 0$. This is the standard symmetric depolarization noise model. Henceforth, we use $p_1$ and $p_2$ to indicate the probability of error for single and two qudit gates respectively.

A noisy gate is modelled as an ideal gate followed by an unwanted Pauli operator \cite{fowler2012surface}. In other words, a one-qubit gate is followed by an unwanted Pauli $\in \{X, Y, Z\}$ with probability $p_x, p_y, p_z$ respectively; and a two-qubit gate is followed by an unwanted Pauli $\in \{I, X, Y, Z\}^{\otimes 2} \setminus \{I,I\}$ with probability $p_i \cdot p_j$, where $i,j \in \{x, y, z\}$. For the sake of convenience, we represent the one-qubit and two-qubit gate error probabilities as $p_1$ and $p_2$ respectively.

In a binary system, there are 2 types each of unwanted $X$ and $Z$ Pauli errors that can follow a one-qubit gate. Therefore, there are $2^2 -1$ ways (without considering the identity error) in which an error can occur after a one-qubit gate. If $p_1$ is the probability of a one-qubit Pauli error, then the evolution of the system under noisy one-qubit operations can be represented as in Eq.~\ref{eq:single_gate_err}.

\begin{align}
    \mathcal{E}(\sigma) = (1-(2^2-1)p_1)\sigma + \sum_{jk \in \{0,1\}^2 \setminus 0*2} p_{jk} K_{jk} \sigma K_{jk}^{\dagger}
\label{eq:single_gate_err}
\end{align}

where $K_{jk}$ represents the various Pauli operators.

Similarly, for two-qubit gates, an unwanted Pauli operator can occur on each of the two qubits after the gate operation. Therefore, there are $2^4-1$ ways (excluding the identity operation on both the qubits) in which a gate can be noisy. If $p_2$ is the probability of two-qubit gate errors, then the evolution of the system under noisy two-qubit operations is represented as in Eq.~\ref{eq:two_gate_err}.

\begin{equation}
%\resizebox{0.97\hsize}{!}{%
    \mathcal{E}(\sigma) = (1-(2^4-1)p_2)\sigma + \sum_{jklm \in \{0,1\}^2 \setminus 0*2} p_{jklm} K_{jklm} \sigma K_{jklm}^{\dagger}
\label{eq:two_gate_err}
\end{equation}

where $p_{jklm} = p_{jk} \cdot p_{lm}$. The probability that the density matrix remains error free is independent of whether the underlying depolarizing channel is symmetric or asymmetric. Rather, it depends on the total probability of error.

Our used decomposition here deals with two-qutrit gates only on ternary
quantum systems. In general, for the decomposition of a Toffoli gate, our used method uses up to 3 dimension. Therefore, for a $3$-dimensional system, the error in our system scales as $\mathcal{O}(3^4)$ as shown in Eq.~\ref{eq:two qudit gate error}. 
% As $(d+2)$-ary quantum systems have been employed for $n$-qudit MCT decomposition, we have a similar form of the two-qudit gate error channel for $(d+2)$-ary quantum systems:

\begin{equation}
%\resizebox{0.98\hsize}{!}{%
    \mathcal{E}(\sigma) = \{1-(3^4-1)p_2\}\sigma + \sum_{\substack{jklm \in \\ \{0,1,2, \dots, 3\}^4 \setminus 0000}} p_{jklm} K_{jklm} \sigma K_{jklm}^{\dagger}
\label{eq:two qudit gate error}
\end{equation}

 In Table~\ref{tab:compgate}, we show the decrease in the probability of no-error for two-qutrit gates due to the usage of higher dimensions for ternary systems. 

\begin{table}[!htb]
    \centering
    \caption{Probability of success of two-qutrit gates due to the usage of higher dimensions}
    \resizebox{8.7cm}{!}{%
    \begin{tabular}{|c|c|c|}
    \hline
    
   Dimension $d$ & without our decomposition & with our decomposition  \\
    \hline
    2  & $1 - 15p_2$  & $1 - 81p_2$ \\
    \hline
    \end{tabular}}
    \label{tab:compgate}
\end{table}

\subsubsection{Idle error:}

In quantum devices, idle errors mainly focus on the relaxation from higher to lower energy levels. Amplitude damping is another name for this. This noise channel takes qutrits to lower states in an irreversible manner. For qubits, the only amplitude damping channel is from $\ket{1}$ to $\ket{0}$, and we denote this damping probability as $\lambda_1$. For qubits, the Kraus operators for amplitude damping are:
\begin{align} K_0 = \begin{pmatrix} 1 & 0 \\ 0 & \sqrt{1 - \lambda_1} \end{pmatrix} \text{\quad and \quad} K_1 = \begin{pmatrix} 0 & \sqrt{\lambda_1} \\ 0 & 0 \end{pmatrix}
\end{align}

 For qutrits, we also model damping from $\ket{2}$ to $\ket{0}$, which occurs with probability $\lambda_{2}$. For qutrits, the Kraus operator for amplitude damping can be modeled as:

%\begin{align}
$$K_0 = \begin{pmatrix} 1 & 0 & 0 \\ 0 & \sqrt{1-\lambda_1} & 0 \\ 0 & 0 & \sqrt{1 - \lambda_2}  \end{pmatrix}
\text{, }$$
%\end{align}
\begin{align}
K_1 = \begin{pmatrix} 0 & \sqrt{\lambda_1} & 0 \\ 0 & 0 & 0 \\ 0 & 0 & 0  \end{pmatrix} \text{\quad and \quad}
K_{2} = \begin{pmatrix} 0 & 0  & \sqrt{\lambda_{2}}\\ 0 & 0 & 0 \\ 0 & 0 & 0 \end{pmatrix}
\end{align}

In each Kraus Operator $K_i$, the value of $\lambda_i \propto exp(-t/T_{1_i})$, where $t$ is the duration of the computation, and $T_{1_i}$ are the relaxation time. We have qubit quantum devices, where $T_{1_1} \simeq 100 \mu s$ in some higher end IBM Quantum Devices \cite{ibmquantum}. However, due to the lack of qudit quantum computers, we do not have explicit values of other $T_{1_i}$'s except $30 \mu s$ for qutrit ($T_{1_2}$) quantum devices \cite{https://doi.org/10.48550/arxiv.2203.07369}. Nevertheless, the length of time depends on the circuit depth. As a result, by reducing the circuit depth, idle error are reduced. Therefore, since depth has been optimized, the decoherence owing to our employed decomposition is significantly lower than the previous decomposition.

\subsection{Analysis of Success Probability}\label{prob_success}

In \cite{Gokhale_2019}, the authors advocated the use of higher dimension for efficient decomposition of Toffoli gates. However, because there was no qudit hardware at the time, they assumed that the value of $T_{1_i}$ for ternary systems is the same as for binary systems in that article. However, as explained in the preceding article, currently we have the value of $T_{1_3}$ for ternary systems. As a result, in this section, we apply the method in \cite{Selinger_2013, Chakrabarti_2021} and our method in this article to investigate the probability of success in the decomposed circuit of a Toffoli gate. It is worth noting that, although the decomposition of \cite{Selinger_2013, Chakrabarti_2021} requires one-qubit and two-qubit gates, our method just necessitates two-qutrit ternary gates.

The complexity of decomposition of a Toffoli gate in terms of the number of gates and depth of the circuit for the method in \cite{Selinger_2013, Chakrabarti_2021} and our used one are depicted here. For each Toffoli decomposition as shown in Fig. \ref{fig:tof_selinger}, one requires 7 one-qubit gates and 16 two-qubit gates with overall circuit depth 7, whether we need 3 two-qutrit gate with circuit depth 3 for our used decomposition. 

Small errors in quantum circuit gates can be described as an ideal gate followed by an undesirable Pauli operator, as discussed in previous subsections. However, instead of comparing the probability of minor errors in the circuit, we compare the probability that the circuit remains error-free (probability of success) as per \cite{majumdar2021optimizing} for the decomposition in \cite{Selinger_2013, Chakrabarti_2021} and our employed decomposition, without losing generality.

For any decomposition, the generalized formula for probability of success ($P_{success}$) described in \cite{majumdar2021optimizing, amitpra} is the product that the individual components does not fail. In other words,
\begin{equation}
%\label{eq:success}
    P_{success} = \Pi_{gates} ({(P_{success~of~ gate})}^{\#~gates} \times  e^{-(depth/T_1)}), 
\end{equation}

where the first term's product is the likelihood of all types of gates employed in the decomposition (one-qubit, two-qubit, two-qutrit), and the second term is the probability of no relaxation error. Due to the zero in the power, when a specific type of gate is not employed in a decomposition, the associated term has a value of $1$. For example, our suggested decomposition does not require a one-qutrit gate, therefore that term's contribution to the product is 1.

Current quantum devices are mostly binary, and the probabilities of one-qubit and two-qubit gates in the IBMQ Quantum Devices are in the range of $10^{-4}$ and $10^{-2}$ respectively \cite{ibmquantum}. Moreover, the time $T_{1_1}$  of most of the IBM Quantum Devices are in the range of $100 \mu s$.  However, in \cite{https://doi.org/10.48550/arxiv.2203.07369}, the authors experimentally showed that the value of $T_{1_2}$  for each ternary gate is $30 \mu s$, which we have also assumed for our study. We assume that the probability of error of each two-qubit and two-qutrit gate is $10^{-2}$, that of one-qubit gate is $10^{-4}$, and the time $T_{1_1}$ is $100 \mu s$ and $T_{1_2}$ is $30 \mu s$ for our simulation.

\begin{figure}
    \centering
    \includegraphics[scale=0.35]{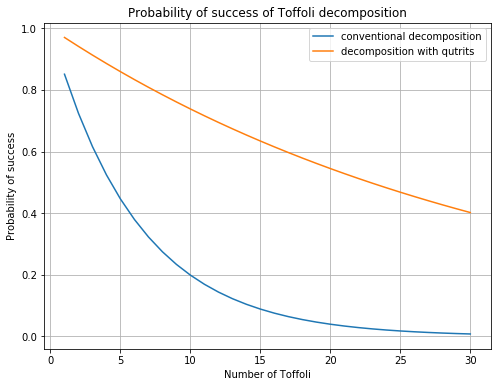}
    \caption{Probability of success for the decomposition of Toffoli gate using our used method versus the method in \cite{Selinger_2013, Chakrabarti_2021}}
    \label{fig:success}
\end{figure}

In Fig.~\ref{fig:success}, we exhibit the probability of success for the Toffoli gate decomposition using the method of \cite{Selinger_2013, Chakrabarti_2021} (which we label as `conventional decomposition') and our used method. We find that our proposed technique produces much fewer errors than the decomposition in \cite{Selinger_2013, Chakrabarti_2021}. This is due to the fact that our decomposition has fewer gates and is shallower. Although our decomposition employs a few qutrit gates, which have a higher error probability due to the curse of dimensionality, our technique is superior due to the overall large reduction in gate count and depth. In fact, for Toffoli count 30 in a circuit, our used decomposition has a probability of success of $\simeq 0.4$, whereas that of \cite{Selinger_2013, Chakrabarti_2021} has a probability of success of $\simeq 0.01$. Therefore, we obtain a percentage decrease in the probability of error by $\simeq 40\%$ for Toffoli count 30. Thus, it can be concluded that the conventional decomposition methods attain $100\%$ error for Toffoli count 30 in a circuit, whereas our used decomposition method yields less erroneous results for Toffoli count 30 and above in a circuit. A comparative study of our Toffoli decomposition used in derivative pricing with some previous works \cite{Chakrabarti_2021, Selinger_2013} is shown in Table \ref{tab:comp_resource}. Our work outperforms them in terms of T-depth, overall circuit depth, overall gate count of the circuit and probability of error.

\begin{table}[htb]
    \centering
    \caption{Comparative analysis between conventional approach \cite{Chakrabarti_2021, Selinger_2013} and our approach with qutrits used in derivative pricing problem for the decomposed circuit of Toffoli gate}
    \resizebox{9cm}{!}{%
    \begin{tabular}{|c|c|c|c|c|c|}
    \hline
         & \# T  & \# overall  & \# two-qutrit  & \# Gate  &  Prob. of error (\%) \\
         & depth & circuit depth & gates & count & for Toffoli count 30\\
         \hline
        Decomposition  & 1 & 7 & 0 & 25 & 99.95\\ 
        of \cite{Chakrabarti_2021, Selinger_2013} & & & & &\\
        
        \hline
        Our Decomposition & 0 & 3 & 3 & 3 & 60\\
        \hline
    \end{tabular}}
    \label{tab:comp_resource}
\end{table}

%\textcolor{red}{add a table/high-level comparison between qubit and qutrit approach for derivative pricing. that should be the concluding discussion}

\section{Conclusion}\label{conclusion}

In this work, we have addressed basic quantum arithmetic operations and estimated the resources using intermediate qutrits which also demonstrate that our approach outperforms current techniques in terms of circuit depth and circuit robustness. Further, as an application, we have shown a novel approach of implementing derivative pricing problem by decomposing Toffoli gate into two-qutrit gates with optimized gate count and depth without using any ancilla qubit. We have given a comparative study on resources for derivative pricing to establish that our approach is better than the existing state-of-the-art one \cite{Chakrabarti_2021}. Finally, we have studied the effect of different error models on this decomposition technique. Our study shows that the few  gates in higher dimensional quantum systems which are used in the used decomposition, are prone to more errors. Nevertheless, as we have obtained improved gate count and circuit depth, leading to low total error probability, the gates can operate with high fidelity as compared to state-of-the art work. Through numerical simulation, as an instance, we have shown that intermediate qutrit-based Toffoli decomposition obtained a percentage increase in the probability of success by $\simeq 40\%$ for Toffoli count 30 in a circuit as compared to conventional way of Toffoli decomposition. These improved quantum arithmetic operations pave the way for more detailed time improvement estimate with intermediate qudit to study the decomposition for other quantum circuits. %\textcolor{red}{give quantitative comparison}

% or


%\ifCLASSOPTIONcompsoc
%\else
%\fi

%There is no conflict of interest.

%\ifCLASSOPTIONcaptionsoff
%  \newpage
%\fi




\bibliographystyle{IEEEtran}
\bibliography{Bibfile}



%\vfill


\end{document}